\documentclass[11pt,twoside]{article}

\usepackage{asp2006}
\usepackage{epsf}
\usepackage{psfig}
\usepackage{lscape}

\begin{document}
\title{3D structure of the outer atmosphere: \\
	combining models and observations}   
\author{S. R\'egnier}   
\affil{School of Mathematics and Statistics, University of St Andrews, St
Andrews, KY16 9SS, United Kingdom}    

\begin{abstract} 
In this review, I focus on the structure and evolution of the coronal magnetic
fields modelled from observations. The development of instruments measuring the
photospheric and chromospheric magnetic fields with a high spatial and time
resolutions allows us to improve the modeling of the coronal fields based on
extrapolation and evolution techniques. In particular, I detail the advance
modelling of quiet-Sun areas, active regions and full-disc evolution. I discuss
the structure of coronal magnetic features such as filaments, sigmoids and
coronal loops as well as their time evolution and instability. The complexity of
the coronal field and the origin of open flux are also investigated in these
different areas. Finally I discuss the future improvements in terms of
instruments and models required to understand better the coronal field.  
\end{abstract}



\section{Introduction}
The structure of the solar corona is organised by its magnetic field. The plasma
$\beta$, the ratio of the plasma pressure to the magnetic pressure, is less than
1 from the bottom of the corona to about 2.5 solar radii \citep{gar01}.
Unfortunately to date, it is not possible to obtain a reliable measurement of
the full magnetic vector in the corona, whilst it is routinely observed in the
photosphere and chromosphere. Waiting for improved spectropolarimetric
observations based on coronal lines, the coronal magnetic field has thus to be
derived from physical assumptions relying on photospheric measurements.

The physical assumptions to model the coronal field depend on the areas of the
Sun considered. Three different parts are often distinguished: (i) the active
regions, (ii) the quiet Sun, and (iii) the full Sun. The active regions on the
photosphere are regions of strong magnetic field which can form sunspots. The
time evolution of active region is 10-15 min by considering an average Alfv\'en
transit time along a loop of 200 Mm (except during flaring activity). The
spatial scale is typically 300-500 Mm. Observations in a broad range of
wavelengths show coronal structures such as filaments/prominences, sigmoids,
loops. Active regions are the sources of eruptions such as flares and Coronal
Mass Ejections (CMEs) due to the large amount of magnetic energy stored. The
quiet Sun is the area outside active regions (including coronal holes). The
typical spatial and time scales are imposed by the structure of granules (20 Mm
and few minutes). The full Sun encompasses both the quiet Sun and active region
areas. At the current spatial resolution, the time of evolution is imposed by
the evolution of active regions for instantaneous magnetogram and by the
differential rotation for synoptic magnetic maps. The models are imposed by the
physical conditions of the different regions and constrained by the
observations.  

In this review, I only focus on the models developed to determine the
different structures of the coronal magnetic field from photospheric
observations with a special emphasize on nonlinear force-free models.  

\section{Force-free models}

	\subsection{Magnetic field extrapolations}

Magnetic field extrapolations consist in computing the coronal magnetic field
assuming an equilibrium state and using the distribution of the magnetic field
observed in the photosphere or chromosphere as boundary condition. This is a
static model of the corona. In the corona, three main forces act on the plasma:
the plasma pressure gradients, the gravitational force and the magnetic forces.
The equation governing the equilibrium is then:
\begin{equation}
- \vec \nabla p + (\vec \nabla \wedge \vec B) \wedge \vec B + \rho \vec g = \vec
0.
\end{equation}

Several main assumptions are thus defined depending on the time and spatial
scales to describe and, most importantly, on the magnetic field measurements
available. It is worth noticing at this stage that none of these assumptions can
describe the real physical nature of the corona as plasma flows have an 
important role. Nevertheless the study of magnetic equilibria remains a key to
understand better the complexity of the coronal magnetic field and, to date,
this is the most reliable method to access the 3D coronal field from
observations.

Three main assumptions are currently in use to extrapolate the magnetic field
into the corona: potential field for which no electric currents (or curl of
magnetic field) are present in the configuration \citep{sch64, sem68}, the
linear force-free field in which the electric currents are parallel to the
magnetic field line and the coefficient of proportionnality is the same
everywhere in the volume \citep{nak72, chi77, ali81, sem88, gar89}, and the
nonlinear force-free field in which the coefficient of proportionnality varies
from one field line to an other \citep[e.g.,][]{wol58, sak81, aly84}. The latter
assumption is the most realistic and most advanced technique in use. Nonlinear
force-free extrapolation techniques can be classified depending on the boundary
conditions they use or their numerical schemes: optimisation  \citep{whe00,
wie04, wie06a, wie06, wie08, tad09}, Grad \& Rubin \citep{gra58, sak81, aly89, ama97,
ama99, whe04, ama06, inh06, whe06, whe07, whe09}, evolutionnary techniques
\citep{mik94}, magneto-frictional \citep{yan86, van00}, vertical integration
\citep{wu90, dem92, son06}, boundary integrals \citep{yan00, yan06, val05}. In
recent reviews \citep{sch06, reg07, wie08a}, the pros and cons of the different
numerical schemes are discussed and compared using semi-analytical models or
observations. To determine nonlinear force-free configurations, the boundary
conditions are the vertical or radial component of the magnetic field, and
either the distribution of $\alpha$ in one polarity or the two transverse
components of the magnetic field in both polarities. The first set of boundary
conditions corresponds to a mathematically well-posed boundary value problem
\citep{gra58, sak81}.   

Recently, more sophisticated assumptions have been developed to improve the
physical content of the above models: the magnetohydrostatic model which takes
into account the plasma pressure gradients and/or the gravitational force
\citep{low85, bog86, low91, neu95, wie07, rua08}, and non force-free models
\citep{hu06, hu08, hu08a, gar09}.

The force-free reconstruction is applied to an observed magnetogram at a given
time and without a priori on the structure of the coronal field. Several other
methods have been developed to construct force-free equilibria adding
constraints from observations. In \cite{van04}, a weakly twisted flux rope is
inserted into a potential field configuration and then relaxed to a nonlinear
force-free state. The flux rope insertion model is constrained by chromospheric
or coronal observations. Unlike the nonlinear force-free reconstructions
mentioned above, the flux rope insertion model only requires the vertical or
radial component of the magnetic field measured on the photosphere. In addition,
a weakly magnetohydrostatic model based on \cite{low91} has been developed by
\cite{aul99} imposing an a priori external bipole allowing the existence of
twisted flux bundle.

	\subsection{Magnetic field evolution}

In order to follow the evolution of the solar corona, two approaches can be
followed. 

First, a time series of equilibria can be constructed from observed magnetograms
assuming that the time of evolution of the coronal structures is slow enough
compared to the reconnection time and the Alfv\'en transit time. The method does
not consider the history of the region as the magnetograms are treated
independently. Nevertheless, part of the history of the region is included in
the electric currents for nonlinear force-free models. The technique of
succesive force-free equilibria as been applied by \cite{hey84} for linear
force-free fields, by \cite{reg06} for nonlinear force-free fields.

Second, the flux transport model is used to describe the long-term evolution of
the solar corona during a magnetic cycle \citep{mac06a, mac06b, yea07, yea08,
yea08a, yea09}. The flux transport model is twofold: (i) evolution of the
photospheric magnetic field, (ii) magneto-frictional relaxation to a nonlinear
force-free equilibrium. The photospheric boundary conditions (usually synoptic
maps) are evolved in time by including in the induction equation the effects of
differential rotation, meridional flow and surface diffusion. Those three
effects have different characteristic times: 0.25 years for the differential
rotation, 2 years for the meridional flow and 34 years for the surface
diffusion. Once the photospheric field is determined, the coronal magnetic field
is derived from the magneto-frictional relaxation method allowing the magnetic
configuration to relax to a nonlinear force-free state. This method takes into
account the history of the magnetic field during a cycle and also includes an
automatic procedure of emerging magnetic bipoles based on the best match with
coincident observations. The flux transport model is used to describe the large
scale structure of the corona as a nonlinear force-free field, and thus gives
physical insights different from the large-scale potential field commonly in
use. As this review focuses on the nonlinear force-free modelling of the solar
corona, I will omit the magnetohydrodynamic (MHD) models of coronal fields.

\section{Structure of the solar corona}

The above techniques to derive the force-free nature of solar corona have been
applied to different magnetic regions: (i) the quiet Sun, (ii) the active
regions, and (iii) the full Sun. 

	\subsection{Quiet Sun}

The quiet Sun is a misnomer. The evolution of the quiet Sun has a characteristic
time of the granule evolution of few minutes. Consequently, lots of eruptive 
events are observed continuously. To model the magnetic field in observed
quiet-Sun regions, the model used is the potential field because the magnetic
field measurements are mostly provided by the line-of-sight component of the
magnetic field. Nevertheless recent development in instrumentation shows the
possibility to measure reliably the three components of the magnetic field with
a great accuracy. 

The quiet-Sun magnetic field, the so-called magnetic carpet, has been modeled as
a potential field defining the polarities as point sources. This is the point
charge method \citep{sch02, lon03, clo03}. These models are based on SOHO/MDI
line-of-sight magnetograms as boundary conditions. SOHO/MDI has a moderate
spatial resolution of 1.98 arcsecond and a time cadence of at most 1 min.
\cite{reg08} have computed the potential field of a quiet-Sun region observed by
Hinode/SOT/NFI with a spatial resolution of 0.16 arcsecond. This model does
consider a continuous distribution of the magnetic field on the photosphere. In
addition, as revealed by previous work, the complexity of the quiet-Sun magnetic
field lies near the photospheric surface (below 5 Mm), therefore \cite{reg08}
have implemented a stretch grid along the vertical axis with a very fine grid
near the bottom boundary in order to resolve the nonlinearities of the magnetic 
field. The authors revealed that the complexity of the magnetic field (defined
as the number of null points) is concentrated in the photosphere and the
chromosphere (below 3.5 Mm) whilst the corona above a quiet-Sun region is not
complex. By measuring the vector magnetic field, Hinode/SOT has successfully
measured the magnetic field in coronal holes, and thus shows the almost unipolar
nature of coronal hole: the small bipoles being connected at low height in the
chromosphere or the bottom of the corona. Consequently, the open magnetic flux
responsible for the fast solar wind has a strong latitudinal dependence as the
quiet-Sun magnetic field becomes more and more unipolar from the equator to
the poles. 

The structure of the coronal field in the quiet Sun is complex and dynamic. The
magnetic field evolves on the time scale of a granule.

	\subsection{Active regions}

The potential field is a minimum of magnetic energy for a given distribution of
the vertical or radial magnetic field component on the photosphere. Therefore
there is no free magnetic energy, no shear and/or twisted field lines in a
potential field configuration. For these reasons, the nonlinear force-free field
is more adequat to describe better the nature of the corona as it contains free
magnetic energy and sheared and twisted flux bundles.

Regarding the magnetic energy, it has been found that an active region contains
enough free magnetic energy to trigger flares \citep[e.g.,][]{reg02, ble02,
reg07a}. By studying the magnetic energy budget before and after a flare, it is
difficult to conclude as often the magnetic energy released during the flare is
in competition with the continuous injection of energy from the convection zone
as well as the redistribution of the energy inside the volume considered
\citep{ble02, reg06, tha08, su09}. 

The magnetic helicity of the magnetic field is a quantity more difficult to
tackle as the knowledge of the vector potential is required inside the coronal
volume. It has been shown that the magnetic helicity is not a conserved quantity
in a volume above an active region as magnetic helicity is injected from the
convection zone and ejected away from the corona. 

In terms of the structure of the corona, the nonlinear force-free field based on
vector magnetograms has revealed the existence of weakly and highly twisted flux
bundles in active regions describing solar features such as:
\begin{itemize}
\item[-]{Filaments:} filaments are magnetic structures containing cool and dense
material compared to the coronal environment. In nonlinear force-free models,
filaments are often identified as weakly twisted flux bundles with magnetic dips
in which the plasma is stored \citep{aul99, yan00, reg04, wie05, dud08, yea09}.
Nevertheless observations have shown that active region filaments can be highly
twisted as they are subject to kink instability; 
\item[-]{Sigmoids:} sigmoids are observed in soft X-ray as S or inverse-S shaped
structures of hot plasma. These structures have been often identified as weakly
or highly twisted flux bundles with no magnetic dips \citep{reg04, can09, su09,
sav09}. As shown by \cite{reg07b}, highly twisted flux tubes are required to
store magnetic energy high in the corona;
\item[-]{Others:} other twisted flux bundles are present in magnetic
configurations with a different amount of twist and/or a different handedness
(as both signs of currents are observed in a polarity) but which cannot be
identified to observed features \citep{reg04}.
\end{itemize}  

Coronal loops in the core of active regions (observed in soft X-rays) can carry
a significant amount  of current \citep{reg04}, whilst large loops on the edge
of active regions (observed in EUV at 1-1.5 MK for instance) are close to
potential magnetic field lines \citep{der09}. 

It is important to notice that the structure of the magnetic field strongly
depends on the magnetic field model and on the nature of the active region,
especially the total magnetic flux and the distribution of polarities on the
photosphere \citep{reg07b}. In particular, \cite{reg07b} demonstrated that,
statistically, the magnetic field lines are higher and longer in a nonlinear
force-free configuration than in a potential field one.

	\subsection{Full Sun}

The nonlinear force-free description of the whole corona is derived from the
flux transport model. Starting from a potential field equilibrium, the
photospheric magnetic distributions are evolved to match the observed synoptic
maps and the 3D coronal field is thus given by a series of nonlinear force-free
equilibria. Note that, compare to previous models, the flux transport model does not
reset the coronal field to a potential field at each time step. As mentioned in
the previous section, statistically speaking, the field lines in a nonlinear
force-free model are longer and higher than in a potential field. The consequence is
that the open magnetic flux contributing to the fast solar wind is larger in
nonlinear force-free models than from potential models. The amount of open flux
from the flux transport model is estimated to be one order of magnitude larger
than for the potential field model \citep{mac10}. The potential models are
useful for a qualitative description of the high corona but improved
models are required to have a better quantitative description of the corona.    

As shown by \cite{coo09}, the complexity of coronal magnetic field is very low:
only few null points (in average, 14 null points) are present in the whole 3D
corona up to 2.5 solar radii. The time variation of the number of null points
follows the magnetic cycle.  

\section{Discussion}	

The nonlinear force-free modelling of the solar corona has became a very
attractive domain of research \citep{sch06, met08, sch08, der09}. This physical
assumption corresponds to an important step in our understanding of the 3D
structure of the solar corona. It corresponds currently to the state-of-the-art
numerical techniques relying on magnetic observations. Nevertheless the
force-free assumption is debatable especially at a time when the new space
missions such as the Hinode satellite have significantly improved the spatial
and time resolutions, and thus show that plasma flows play an important role in
the nature of photospheric and chromospheric plasmas. The next step to improve
this type of modelling based on observations is to consider the plasma
parameter: the magnetohydrostatic model is a step forward to be implemented for
future solar missions. As mentioned already, some tentatives to model the solar
corona as a magnetohydrostatic equilibrium have attempted \citep{wie07}. 

The nonlinear force-free models are constrained by the photospheric or
chromospheric magnetic field. With the large amount of data from Hinode or Solar
Dynamics Observatory (SDO), it is suggested that more constraints should be
taken into account to retrieved a more realistic description of the coronal
field. Attempts have been made in several papers mentioned above
\citep[e.g.,][]{van04, der09}.  

\acknowledgements 
I would like to thank the organisers of the 3rd Hinode meeting in
Tokyo (Japan) for the opportunity to give this review talk. I also thank
Duncan Mackay for fruitful discussions on this topic. I would like to thank the
UK STFC for financial support (STFC RG in St Andrews) and the European
Commission through the SOLAIRE network (MTRN-CT-2006-035484).



\end{document}